\documentclass[a4paper,11pt]{article}
\usepackage{amsmath,amssymb}
\usepackage[numbers,square]{natbib}
\usepackage{amsfonts}
\usepackage{graphicx}

\title{Universal procedure to cure future singularities \\ of dark energy models}

\author{A.~J. L\'opez-Revelles and E. Elizalde \\ \mbox{} \\ 
\small Instituto de Ciencias del Espacio (ICE/CSIC) \, and \\ \small Institut d'Estudis
Espacials de Catalunya (IEEC) \\ \small
Campus UAB, Facultat de Ci\`encies, Torre C5-Parell-2a planta \\ \small E-08193 Bellaterra (Barcelona) Spain \\ \small E-mail: alopez@ieec.uab.es, elizalde@ieec.uab.es}

\date{}

\begin{document}

\maketitle

\begin{abstract}

A systematic search for different viable models of the dark energy universe, all of which give rise to finite-time, future singularities, is undertaken, with the purpose to try to find a solution to this common problem. After some work, a universal procedure to cure all future singularities is developed and carefully tested with the help of explicit examples corresponding to each one of the four different types of possible singularities, as classified in the literature. The cases of a fluid with an equation of state which depends on some parameter, of modified gravity non-minimally coupled to a matter Lagrangian, of non-local gravity, and of isotropic turbulence in a dark fluid universe theory are investigated in detail.

\end{abstract}


\bigskip

	\section{Introduction}

		One of the most important problems of modern cosmology is to explain in a plausible way the late-time acceleration of the universe expansion. Recent observational data, coming from more and more precise high-redshift surveys of type Ia supernovae \cite{Riess,Perlmutter} and from anisotropies in the power spectrum of the cosmic microwave background \cite{Bennett,Netterfield,Halverson}, indicate that, right now, our Universe is undergoing a phase of accelerated expansion. The favored explanation for this behavior is that the universe is filled with some form of dark energy (DE). However, none of the DE models proposed so far is completely satisfactory. In order to explain the late-time cosmic acceleration, and to unify it with inflation, a new family of theories has been proposed, based on modifications of Einstein's gravity (see \cite{N-O.2} for a recent review). These theories, such as $f(R)$ and $f(G)$ modified gravities ($R$ for the curvature and $G$ for the Gauss-Bonnet invariant), and also non-local gravity, have the power to unify primordial inflation with late-time inflation in a natural way. While a number of the original models had to be rejected by several reasons, some subfamilies of them (see \cite{C-E-N-O-S-Z}) fulfill all available cosmological constraints and have been checked to pass the solar system tests and cosmological bounds, e.g., the fact that, at these scales, Einstein's gravity is valid to a very high degree of accuracy. But even these, so called viable, modified gravities are not free from other problems, one of the most important being the frequent appearance of finite-time, future singularities.

		The singularity problem is indeed of fundamental importance in modern cosmology. In order to address this issue rigorously it would be necessary to develop a fully-fledged theory of quantum gravity, but this has proven to be a very difficult, up to now impossible, challenge. Anyhow, the presence of a finite-time, future singularity may cause various problems of physical nature, as instabilities in current black hole and stellar astrophysics. It turns out that, even without the support of a quantum gravity theory, it is still meaningful at first instance (and of great importance) to try to find natural scenarios, already at the classical and semiclassical levels, that may cure the emergence of these finite-time, future singularities.
		Usually, one starts with some given theory and then solves the corresponding equations of motion in order to define the associated background dynamics. But, for several models of modified gravity or scalar-tensor gravity, there is in fact another possibility owing to the fact that those models are defined in terms of some arbitrary functions or potentials. The new possibility consists in using the freedom in the choice of such arbitrary functions or potentials, with the aim to reconstruct the---in general complicated---background cosmology which complies with the latest observational data.

		In this paper, a reconstruction program of this sort is applied to a number of theories which do give rise to models that exhibit finite-time, future singularities, with the purpose to investigate its structure in detail and try then to cure this common problem. Specifically, it will be found in all cases that the addition of an $R^2$ term provides a universal tool capable to cure these finite-time future singularities.

The work is organized as follows.
		In Sect.~2 the case of a fluid with an equation of state (EoS) depending on a parameter, $\alpha$, which can give rise to finite-time future singularities, is considered. Different possibilities for the evolution of this universe, depending on the value of $\alpha$, were studied in \cite{N-O.1}. In the subsections it will be explicitly shown that, adding a function $G(H, \dot H, \ddot H, ...)$ to the EoS of the fluid, the different singularities can indeed be cured. Such function $G$ is actually to be considered as a modification of Einsteinian gravity.
		Sect.~3 is devoted to the case of non-minimal coupling of modified gravity to a matter Lagrangian. The reconstruction scheme is developed for this case and the specific example of the cosmology given by the Hubble function $H(t) = h_s/(t_s - t)$ is analyzed. The calculation of the Friedmann equations for this example of modified gravity non-minimally coupled to the matter Lagrangian is explicitly carried out in App.~A.
		In Sect.~4 the case of non-local gravity is discussed. The example of the de Sitter space is reproduced in the framework of non-local gravity. It is pointed out there that theories of this kind can give rise to a finite-time, future singularity.
		Finally, Sect.~5 is devoted to the case of isotropic turbulence in the dark fluid universe. It will be shown there that the contribution of the turbulent part of dark energy can be reproduced through the use of a scalar-tensor theory. Several examples are discussed in detail. The paper ends with some conclusions and an outlook.

	\section{Accelerating universe with and without a future singularity}

		Different accelerating universes, with and without finite-time future singularities, are considered in this section. We work with a particular fluid with EoS given by:
			\begin{equation}\label{a1}
				p = - \rho + A \rho^\alpha,
			\end{equation}
 The nature of each singularity depends on the value of the parameter $\alpha$. All  possibilities, corresponding to the different values of $\alpha$, have been studied in \cite{N-O.1}. It will be shown below that introducing a specific function, $G(H)$, into Eq.~(\ref{a1}),
			\begin{equation}\label{a2}
				p = - \rho + A \rho^\alpha + G(H),
			\end{equation}
	the	singularity can be avoided.

		The spatially flat Friedmann-Robertson-Walker (FRW) universe in the frame of General Relativity will be considered
			\begin{equation}\label{aa1}
				ds^2 = - dt^2 + a(t)^2 \left( dx^2 + dy^2 + dz^2 \right),
			\end{equation}
		 $a(t)$ being the scale factor. The Friedmann equations are
			\begin{equation}\label{aa2}
				H^2 = \frac{\kappa^2}{3} \rho,
			\end{equation}
			\begin{equation}\label{aa3}
				\dot H + H^2 = - \frac{\kappa^2}{6} (\rho + 3 p),
			\end{equation}
		with $\kappa^2 = 8 \pi G$.
		For future use, it is useful to classify the future singularities as in \cite{N-O-T}, namely,
			\begin{itemize}
				\item  Type I (``Big Rip'') : For $t \to t_s$, $a \to \infty$,
					$\rho \to \infty$ and $|p| \to \infty$. This type of singularity is discussed in \cite{Caldwell:2003vq}.
				\item  Type II (``sudden'') : For $t \to t_s$, $a \to a_s$,
					$\rho \to \rho_s$ and $|p| \to \infty$.
				\item  Type III : For $t \to t_s$, $a \to a_s$,
					$\rho \to \infty$ and $|p| \to \infty$.
				\item  Type IV : For $t \to t_s$, $a \to a_s$,
					$\rho \to \rho_s$, $|p| \to p_s$ and higher derivatives of $H$ diverge.
			\end{itemize}
		Here $t_s$, $a_s$, $\rho_s$ and $p_s$ are constants, with $a_s \neq 0$.

		In the next subsections we will consider four different cases for the dark fluid (\ref{a1}), which lead to the four possible different singularities. We will produce a particular function $G(H)$ for each case, which will cure each specific singularity. Finally, in the last subsection, a general function $G(H, \dot H, \ddot H, ...)$ that cures all possible finite-time future singularities of (\ref{a1}) will be constructed.

		\subsection{$p = - \rho + A \rho^2$}

			For this particular EoS a Type III singularity occurs $(H(t) \propto \left( t_0 - t \right)^{\frac{1}{1 - 2 \alpha}},\ \alpha > 1)$. However, if the following EoS is considered
				\begin{equation}\label{a3}
					p = - \rho + A \rho^2 + G(H),
				\end{equation}
			with
				\begin{equation}\label{a4}
					G(H) = - \frac{9 A}{\kappa^4} H^4 + C,
				\end{equation}
			being $C$ real, then the singularity is avoided. In order to explain this fact, one must take into account that, using the Friedmann equation (\ref{aa2}), Eq.~(\ref{a4}) reduces to
				\begin{equation}\label{a6}
					G(H) = - A \rho^2 + C,
				\end{equation}
			and using now Eq.~(\ref{a6}) the EoS (\ref{a3}) yields
				\begin{equation}\label{a7}
					p = -\rho + C,
				\end{equation}
			which does not have future finite-time singularities of any kind.
			It is interesting to note that the above specific choice of $G(H)$ can be motivated by modified gravity (see \cite{N-O.2}).

		\subsection{$p = - \rho + A \rho^{\frac{2}{3}}$}

			In this case, a Type I singularity occurs, namely $H(t) = \frac{-\frac{2}{3 A}}{t_0 - t},\ \alpha = 1 \ \mbox{and} \ A < 0 \ \ \mbox{or} \ \ H(t) \propto \left( t_0 - t \right)^{\frac{1}{1 - 2 \alpha}},\ 1/2 < \alpha < 1$. Considering now Eq.~(\ref{a3}), with
				\begin{equation}\label{a8}
					G(H) = B H^2,
				\end{equation}
		and	using the Friedmann equation given by (\ref{aa1}), the EoS (\ref{a3}) reduces to
				\begin{equation}\label{a9}
					p = - \rho + A \rho^{\frac{2}{3}} + B' \rho,
				\end{equation}
			where $B' = B \frac{\kappa^2}{3}$. When $\rho \rightarrow \infty$ the term $B' \rho$ dominates over the term $A \rho^{\frac{2}{3}}$ which was the one that caused the singularity. If $B > 0$ then this Type I singularity is removed.

		\subsection{$p = - \rho + A \rho^{\frac{1}{5}}$}

			With this EoS there appears a Type IV singularity, namely $H(t) \propto \left( t_0 - t \right)^{\frac{1}{1 - 2 \alpha}} \ \mbox{and} \ \frac{1}{1 - 2 \alpha} \ \mbox{is not an integer}$. Now, if  Eq.~(\ref{a3}) is considered, with
				\begin{equation}\label{a10}
					G(H) = C,
				\end{equation}
			where $C$ is real, then the EoS (\ref{a3}) turns into
				\begin{equation}\label{a11}
					p = - \rho + A \rho^{\frac{1}{5}} + C.
				\end{equation}
			Whenever $\rho \rightarrow 0$ the term $C$ dominates over the term $A \rho^{\frac{1}{5}}$ and the existing singularity, of Type IV, is cured.

		\subsection{$p = - \rho + A \rho^{-2}$}

			In this case a Type II singularity shows up, namely $H(t) \propto \left( t_0 - t \right)^{\frac{1}{1 - 2 \alpha}},\ \alpha < 0$. By taking in Eq.~(\ref{a3})
				\begin{equation}\label{a12}
					G(H) = - A \frac{\kappa^4}{9} H^{-4} + C,
				\end{equation}
			with $C$ real, and using Friedmann's equation (\ref{aa1}), Eq.~(\ref{a12}) yields
				\begin{equation}\label{a13}
					G(H) = - A \rho^{-2} + C.
				\end{equation}
			Then, the EoS (\ref{a3}) reduces to
				\begin{equation}\label{a13b}
					p = - \rho + C,
				\end{equation}
			which does no more exhibit any kind of finite-time future singularity.

		Summing up, we have shown in all previous situations that, for each case, a particular function, $G(H)$, can be found which cures the singularity which can possibly appear in the model given by Eq.~(\ref{a1}). It is interesting to realize that a function $G(H)$ of this kind can be interpreted as a contribution of modified gravity, as was shown in \cite{N-O.2,B-N-O}. Putting everything together, we thus have demonstrated, in a very explicit way, how modified gravity is able to cure all finite-time future singularities that can possibly appear in a fluid with the particular EoS given by Eq.~(\ref{a1}).

		\subsection{A generic function $G(H, \dot H, \ddot H,...)$ which cures all singularities for the fluid with $p = - \rho + A \rho^\alpha$}

			In this subsection we will develop a systematic method for finding a function $G(H, \dot H, \ddot H,...)$ which avoids any possible singularity for the model with $p = - \rho + A \rho^\alpha$. Let us recall \cite{B-N-O} that every $F(R)$-modified gravity can be seen as Einsteinian gravity with a particular EoS which absorbs the effects of $F(R)$.

			We consider the case
				\begin{equation}\label{a14}
					F(R) = R + f(R),
				\end{equation}
			being $f(R) = a R^2$ (it is known that the term $R^2$ cures the possible appearance of all future finite-time singularities \cite{N-O.2,A-N-O,C-L-N-O}). For Eq.~(\ref{a14}), using the results obtained in \cite{B-N-O}, it follows that
				\begin{equation}\label{a15}
					\rho_{eff} = \frac{1}{\kappa^2} \left[ - \frac{1}{2} f(R) + 3 \left( H^2 + \dot H \right) f'(R) - 18 \left( 4 H^2 \dot H + H \ddot H \right) f''(R) \right] + \rho_{matter},
				\end{equation}
				\begin{equation}\label{a16}
					p_{eff} = \frac{1}{\kappa^2} \left[ \frac{1}{2} f(R) - \left( 3 H^2 + \dot H \right) f'(R) + 6 \left( 8 H^2 \dot H + 4 \dot H^2 + 6 H \ddot H + \dddot H \right) f''(R) + \right.$$
					$$\left. + 36 \left( 4 H \dot H + \ddot H \right)^2 f'''(R) \right] + p_{matter}.
				\end{equation}
			It is also known that the Friedmann equations can be written as
				\begin{equation}\label{a17}
					\rho_{eff} = \frac{3 H^2}{\kappa^2},
				\end{equation}
				\begin{equation}\label{a18}
					p_{eff} = - \frac{1}{\kappa^2} \left( 2 \dot H + 3 H^2 \right).
				\end{equation}
			In the case of $f(R) = a R^2$ and taking into account Eqs.~(\ref{a17}) and (\ref{a18}), Eqs.~(\ref{a15}) and (\ref{a16}) reduce to:
				\begin{equation}\label{a19}
					\rho_{matter} = \rho_{eff} - \frac{18 a}{\kappa^2}  \left( \dot H^2 - 6 H^2 \dot H - 2 H \ddot H \right),
				\end{equation}
				\begin{equation}\label{a20}
					p_{matter} = p_{eff} - \frac{6 a}{\kappa^2}  \left( 9 \dot H^2 + 18 H^2 \dot H + 12 H \ddot H + 2 \dddot H \right),
				\end{equation}
respectively.			
			If we now consider the EoS
				\begin{equation}\label{a21}
					p_{matter} = - \rho_{matter} + A \rho_{matter}^\alpha,
				\end{equation}
			introducing into Eq.~(\ref{a21}), the results obtained in Eqs.~(\ref{a19}) and (\ref{a20}), we get
				\begin{equation}\label{a22}
					p_{eff} = - \rho_{eff} + A \rho_{eff}^\alpha + G(H, \dot H, ...),
				\end{equation}
			where $G(H, \dot H,...)$, in the case of $F(R) = R + a R^2$, is given by:
				\begin{equation}\label{a23}
					G(H, \dot H, ...) = \frac{12 a}{\kappa^2} \left( 6 \dot H^2 + 3 H \ddot H + \dddot H \right) +  \frac{A}{\kappa^{2 \alpha}} \left\{ \left[ 3 H^2 + 18 a \left( \dot H^2 - 6 H^2 \dot H - 2 H \ddot H \right) \right]^\alpha - \left( 3 H^2 \right)^\alpha \right\}.
				\end{equation}
			Thus, using $F(R)$ modified gravity (adding in the action a term proportional to $R^2$, see \cite{N-O.2,A-N-O,C-L-N-O}), a function $G(H, \dot H, ...)$ has been found which cures all the singularities which appeared in the model given by Eq.~(\ref{a21}), in the frame of Einstein's gravity.

	\section{$f(R)$ Modified gravity with possible future singularities for the case: $\mathcal{L} = \frac{1}{\kappa^2} R + f(R) \mathcal{L}_m$}

		In this section we will investigate $f(R)$ modified gravities non-minimally coupled to matter-like Lagrangians that lead to future finite-time singularities. A general Lagrangian density of this sort is (see \cite{N-O.2,N-O.3,A-B-F-O})
			\begin{equation}\label{0}
				\mathcal{L} = \frac{1}{\kappa^2} R + f(R) \mathcal{L}_m.
			\end{equation}
		By varying with respect to the metric, the following field equations are obtained
			\begin{equation}\label{1}
				\frac{1}{\kappa^2} \left( R_{\mu \nu} - \frac{1}{2} g_{\mu \nu} R \right) + R_{\mu \nu} f'(R) \mathcal{L}_m + \left( g_{\mu \nu} \nabla^2 - \nabla_\mu \nabla_\nu \right) \left( f'(R) \mathcal{L}_m \right) - \frac{1}{2} f(R) T_{\mu \nu} = 0.
			\end{equation}
		Furthermore, considering the particular case given by $\mathcal{L}_m = \partial^\mu \phi \partial_\mu \phi$, and varying with respect to the field, we get
			\begin{equation}\label{2}
				\partial_\mu \left( \sqrt{-g} f(R) \partial^\mu \phi \right) = 0,
			\end{equation}
	and, if it is now assumed that $\phi = \phi(t)$, Eq.~(\ref{2}) reduces to
			\begin{equation}\label{3}
				\partial_t \left( \sqrt{-g} f(R) \partial^t \phi \right) = 0 \Rightarrow \sqrt{-g} f(R) \partial_t \phi = 0.
			\end{equation}
		Considering a spatially flat FRW universe, this is $\sqrt{-g} = a(t)^3$, and writing $\partial_t \phi = \dot \phi$, Eq.~(\ref{3}) yields
			\begin{equation}\label{4}
				a(t)^3 f(R) \dot \phi = C,
			\end{equation}
		where $C$ is a constant. Then,
			\begin{equation}\label{5}
				\dot \phi = \frac{C}{a(t)^3 f(R)}.
			\end{equation}
		Taking once more into account that $\phi = \phi(t)$, one gets
			\begin{equation}\label{6}
				\mathcal{L}_m = - \dot \phi^2 = - \frac{C^2}{a(t)^6 f(R)^2}.
			\end{equation}
		Thus, considering a spatially-flat FRW universe and Eq.~(\ref{6}), Friedmann's equations for the Lagrangian density given by (\ref{0}) can be derived. Details of the long calculations leading to these equations are given in appendix \ref{appendixA}. One should note that (\ref{14}) and (\ref{15}) constitute a pair of differential equations for $f(R)$ with the variable being the scalar curvature, $R$. Thus, starting from a given Hubble function $H(t)$ and taking into account that $t = t(R)$, from the relation $R = 6 \dot H(t) + 12 H(t)^2$ and by using (\ref{14}) or (\ref{15}) we obtain a specific function $f(R)$ that reproduces the given Hubble function.

		Another important remark is the following. From the stress-energy tensor $T_{\mu \nu}$
			\begin{equation}\label{7}
				T_{\mu \nu} = - \frac{2}{\sqrt{-g}} \frac{\delta \left( \sqrt{-g} \mathcal{L}_m \right)}{\delta g^{\mu \nu}} \Rightarrow \left\{ \begin{array}{l}
						T_{00} = - \dot \phi^2 = - \frac{C^2}{a(t)^6 f(R)^2} \\
						T_{ii} = - a(t)^2 \dot \phi^2 = - \frac{C^2}{a(t)^4 f(R)^2}
					\end{array}
				\right. ,
			\end{equation}
by comparison of
			\begin{equation}\label{16}
				T_{\mu \nu} = - \frac{2}{\sqrt{-g}} \frac{\delta \left( \sqrt{-g} \mathcal{L}_m \right)}{\delta g^{\mu \nu}} \Rightarrow \left\{ \begin{array}{l}
						T_{00} = - \rho g_{00} = \rho \\
						T_{ii} = p g_{ii} = a(t)^2 p
					\end{array}
				\right.
			\end{equation}
		with (\ref{7}), we obtain the result:
			\begin{equation}\label{17}
				p = \rho = - \dot \phi^2 = - \frac{C^2}{a(t)^6 f(R)^2}.
			\end{equation}
		Now, from (\ref{17}) we know that, for this model, a singularity of type II is avoided (for the classification of the future singularities, see \cite{N-O-T} or Sect.~II above).

		In order to find a model with the Lagrangian density (\ref{0}) that ends in a future finite-time singularity, we now consider the case:
			\begin{equation}\label{18}
				H(t) = \frac{h_s}{t_s - t}
			\end{equation}
	where, from $R = 6 \dot H(t) + 12 H(t)^2$, it follows that
			\begin{equation}\label{19}
				t = t_s - \frac{h_1}{\sqrt{R}},
			\end{equation}
		where $h_1 = \sqrt{6 h_s (1 + 2 h_s)}$. Then, we can write
			\begin{equation}\label{20}
				H = h_2 R^{\frac{1}{2}},
			\end{equation}
			\begin{equation}\label{21}
				\dot H = h_3 R^{\frac{2}{2}},
			\end{equation}
			\begin{equation}\label{22}
				\ddot H = h_4 R^{\frac{3}{2}},
			\end{equation}
		being $h_2 = \sqrt{\frac{h_s}{6 (1 + 2 h_s)}}$, $h_3 = \frac{1}{6 (1 + 2 h_s)}$ and $h_4 = \frac{1}{3 (1 + 2 h_s) \sqrt{6 h_s (1 + 2 h_s)}}$. For (\ref{18}), we have
			\begin{equation}\label{23}
				\int \limits_{t_0}^{t(R)} H(t') dt' = \ln{\left( \frac{t_s - t_0}{h_1} \sqrt{R} \right)^{h_s}},
			\end{equation}
		hence, Eq.~(\ref{14}) reduces to
			\begin{equation}\label{24}
				f(R) + a_1 R^{a_2} f(R)^2 + a_3 R \frac{df(R)}{dR} + 2 a_4 \frac{R^2}{f(R)} \left( \frac{df(R)}{dR} \right)^2 - a_4 R^2 \frac{d^2 f(R)}{dR^2} = 0,
			\end{equation}
		where $a_1 = \frac{a_0^6 h_2^2}{\kappa^2 C^2} \left( \frac{t_s - t_0}{h_1} \right)^{6 h_s}$, $a_2 = 3 h_s + 1$, $a_3 = h_3 + 7 h_2^2$ and $a_4 = 6 h_2 (h_4 + 4 h_2 h_3)$. The solution for Eq.~(\ref{24}) gives us the function $f(R)$ that reproduces the Hubble function given by (\ref{18}). Eq.~(\ref{24}) can be solved, but the solution obtained is too long and not particularly insightful to be written here.

		To conclude, we should mention that quantum gravity effects (which usually contain different powers of the curvature) become very important near the future singularity (see \cite{E-N-O}). There, classical considerations are not valid. It is known \cite{N-O.2,A-N-O,C-L-N-O} that the $\Box R$ term works against the singularity. Thus, an $R^2$ term (which will generate a $\Box R$ term) would cure the possible singularities that could arise in the theory.

	\section{Future finite-time singularities in non-local gravity}
		
		The case of non-local gravity will be here considered. This theory gives a natural unification of inflation with the current cosmic acceleration and it is inspired by quantum loop corrections (see \cite{N-O.2,D-W,N-O.4,J-N-O-S-T-Z}).
		Non-local effects come from the introduction in the action of the inverse of the D'Alembertian, $\Box^{-1}$, and the simplest action of non-local gravity is therefore
			\begin{equation}\label{nlg1}
				S = \int d^4 x \, \sqrt{-g} \, \left\{ \frac{1}{2 \kappa^2} \left[ R \left( 1 + f(\Box^{-1} R) \right) - 2 \Lambda \right] + \mathcal{L}_{matter}(Q;g) \right\},
			\end{equation}
		where $Q$ stands for the matter fields and $\Lambda$ is the cosmological constant. Introducing two scalar fields, $\eta$ and $\xi$, action (\ref{nlg1}) can be rewritten as
			\begin{equation}\label{nlg2}
				S = \int d^4 x \, \sqrt{-g} \, \left\{ \frac{1}{2 \kappa^2} \left[ R \left( 1 + f(\eta) \right) + \xi \left( \Box \eta - R \right) - 2 \Lambda \right] + \mathcal{L}_{matter} \right\}
			\end{equation}
		If we assume a spatially-flat FRW metric, and that $\eta = \eta(t)$ and $\xi = \xi(t)$, the equations of motion for the scalar fields and the Friedmann equations read (see \cite{N-O.2})
			\begin{equation}\label{nlg3}
				0 = \ddot \eta + 3 H \dot \eta + 6 \dot H + 12 H^2,
			\end{equation}
			\begin{equation}\label{nlg4}
				0 = \ddot \xi + 3 H \xi - (6 \dot H + 12 H^2) f'(\eta),
			\end{equation}
			\begin{equation}\label{nlg5}
				0 = - 3 H^2 \left( 1 + f(\eta) - \xi \right) + \frac{1}{2} \dot \xi \dot \eta - 3 H \left( f'(\eta) \dot \eta - \dot \xi \right) + \Lambda + \kappa^2 \rho_{matter},
			\end{equation}
			\begin{equation}\label{nlg6}
				0 = \left( 2 \dot H + 3 H^2 \right) \left( 1 + f(\eta) - \xi \right) + \frac{1}{2} \dot \xi \dot \eta + \left( \frac{d^2}{dt^2} + 2 H \frac{d}{dt} \right) \left( f(\eta) - \xi \right) - \Lambda + \kappa^2 p_{matter}.
			\end{equation}
		Given a Hubble function, $H(t)$, Eq.~(\ref{nlg3}) can then be solved to obtain $\eta = \eta(t)$; moreover, the function $\xi = \xi(t)$ can be obtained from Eq.~(\ref{nlg4}) if one assumes a form for the function $f(\eta)$. Once we have the functions $\eta(t)$ and $\xi(t)$, assuming an EoS for $p_{matter}$ and $\rho_{matter}$, Eqs.~(\ref{nlg5}) and (\ref{nlg6}) yield the relation between the different parameters that appear in the model (i.e. the constants of integration of the functions $\eta(t)$ and $\xi(t)$, the cosmological constant $\Lambda$, etc.).
		
		In \cite{N-O.2} the previous scheme is used to show that de Sitter space ($H(t) = H_0$) can be a solution in non-local gravity. This is the case for matter of constant EoS $\omega$, when
			\begin{equation}\label{nlg7}
				H = H_0,
			\end{equation}
			\begin{equation}\label{nlg8}
				\eta (t) = - 4 H_0 t,
			\end{equation}
			\begin{equation}\label{nlg9}
				f(\eta) = f_0 e^{\frac{\eta}{\beta}} = f_0 e^{- \frac{4 H_0 t}{\beta}},
			\end{equation}
			\begin{equation}\label{nlg10}
				\xi(t) = - \frac{3 f_0 \beta}{3 \beta - 4} e^{- \frac{4 H_0 t}{\beta}} - \xi_1,
			\end{equation}
		with $H_0$, $f_0$, $\beta$ and $\xi_1$ being constants. For de Sitter space and for matter with constant EoS $\omega$, the energy density is
			\begin{equation}\label{nlg10b}
				\rho _{matter}(t) = \rho_0 e^{- 3 (1 + \omega) H_0 t}.
			\end{equation}
		In order to fulfill Eqs.~(\ref{nlg3}), (\ref{nlg4}), (\ref{nlg5}) and (\ref{nlg6}), it is necessary that
			\begin{equation}\label{nlg11}
				\beta = \frac{4}{3(1 + \omega)},
			\end{equation}
			\begin{equation}\label{nlg12}
				f_0 = \frac{\kappa^2 \rho_0}{3 H_0^2 (1 + 3 \omega)},
			\end{equation}
			\begin{equation}\label{nlg13}
				\xi_1 = -1 + \frac{\Lambda}{3 H_0^2}.
			\end{equation}
		Thus, de Sitter space can indeed be a solution of non-local gravity.

		There are also singular solutions; however, they are very involved and will not be considered here. These singular solutions could also be cured by the addition of an $R^2$ term (inspired by quantum gravity effects near the singularity), a procedure which could again turn into a universal tool in order to suppress all finite-time future singularities, as before (see \cite{N-O.2,A-N-O,C-L-N-O}).

	\section{Reproducing isotropic turbulence in a dark fluid universe with scalar-tensor gravity}
		
		We will now emphasize the fact that a scalar-tensor theory can be used \cite{C-N-O-T} in order to reproduce isotropic turbulence in a dark fluid universe  \cite{B-G-N-O}. To this end, let us consider the following scalar-tensor theory action
			\begin{equation}\label{it1}
				S = \int d^4x \, \sqrt{-g} \, \left[ \frac{1}{2 \kappa^2} R \, - \, \frac{1}{2} \omega(\phi) \partial_{\mu} \phi \partial^{\mu} \phi \, - \, V(\phi) \, + \, \mathcal{L}_{matter} \right],
			\end{equation}
		which leads to the Friedmann equations
			\begin{equation}\label{it2}
				\frac{3 H^2}{\kappa^2} = \frac{1}{2} \omega(\phi) \dot \phi^2 \, + \, V(\phi) \, + \, \rho_{matter},
			\end{equation}
			\begin{equation}\label{it3}
				- \frac{2 \dot H + 3 H^2}{\kappa^2} = \frac{1}{2} \omega(\phi) \dot \phi^2 \, - \, V(\phi) \, + \, p_{matter}.
			\end{equation}
		If we consider that the scalar part of the action dominates over the matter part, these Friedmann equations reduce to
			\begin{equation}\label{it4}
				\frac{3 H^2}{\kappa^2} = \frac{1}{2} \omega(\phi) \dot \phi^2 \, + \, V(\phi),
			\end{equation}
			\begin{equation}\label{it5}
				- \frac{2 \dot H + 3 H^2}{\kappa^2} = \frac{1}{2} \omega(\phi) \dot \phi^2 \, - \, V(\phi).
			\end{equation}
		On the other hand, the Friedmann equations describing isotropic turbulence in a dark fluid universe (see \cite{B-G-N-O}) are
			\begin{equation}\label{it6}
				\frac{3 H^2}{\kappa^2} = \rho_{dark} \, + \, \rho_{turb} \, + \, \rho_{rad} \, + \, \rho_{matter},
			\end{equation}
			\begin{equation}\label{it7}
				- \frac{2 \dot H + 3 H^2}{\kappa^2} = p_{dark} \, + \, p_{turb} \, + \, p_{rad} \, + \, p_{matter}.
			\end{equation}
		Now, under the proviso that the turbulent part $\rho_{turb}$ dominates, Eqs.~(\ref{it6}) and (\ref{it7}) reduce to, respectively,
			\begin{equation}\label{it8}
				\frac{3 H^2}{\kappa^2} = \rho_{turb},
			\end{equation}
			\begin{equation}\label{it9}
				- \frac{2 \dot H + 3 H^2}{\kappa^2} = p_{turb}.
			\end{equation}

		Thus, comparing Eqs.~(\ref{it4}) and (\ref{it5}) with Eqs.~(\ref{it8}) and (\ref{it9}),
			\begin{equation}\label{it10}
				\rho_{turb} = \frac{1}{2} \omega(\phi) \dot \phi^2 \, + \, V(\phi),
			\end{equation}
			\begin{equation}\label{it11}
				p_{turb} = \frac{1}{2} \omega(\phi) \dot \phi^2 \, - \, V(\phi),
			\end{equation}
		and assuming at this point that $\phi = t$, we get the expressions
			\begin{equation}\label{it12}
				\rho_{turb} = \frac{1}{2} \omega(\phi) \, + \, V(\phi),
			\end{equation}
			\begin{equation}\label{it13}
				p_{turb} = \frac{1}{2} \omega(\phi) \, - \, V(\phi),
			\end{equation}
		respectively. Finally, the functions $\omega(\phi)$ and $V(\phi)$ are given by
			\begin{equation}\label{it14}
				\omega(\phi) = \rho_{turb} \, + \, p_{turb},
			\end{equation}
			\begin{equation}\label{it15}
				V(\phi) = \rho_{turb} \, - \, p_{turb}.
			\end{equation}
			
		In \cite{B-G-N-O} some examples of isotropic turbulence were given. In the following subsections, by using Eqs.~(\ref{it14}) and (\ref{it15}), scalar-tensor gravities that reproduce these specific examples will be constructed. In what follows below, the EoS $p_{turb} = \omega_{turb} \rho_{turb}$ is assumed for the isotropic turbulence.
		
		\subsection{De Sitter space}
			
			In this subsection we consider the case of de Sitter space, for which the scale factor is given by $a(t) = a_0 e^{H_0 t}$, with constant $a_0$ and $H_0$. For this case, the energy density of the turbulent part is
				\begin{equation}\label{it16}
					\rho_{turb} = e^{- 3 \gamma_{turb} H_0 t} \, \left[ \frac{C}{3 \gamma_{turb}} \left( e^{- \frac{5}{2} \gamma_{turb} H_0 t_{in}} - e^{- \frac{5}{2} \gamma_{turb} H_0 t} \right) \, + \, \frac{2 C_0 a_0^{\frac{5}{2} \gamma_{turb} H_0}}{5 \gamma_{turb}} \right]^{- \frac{6}{5}},
				\end{equation}	
			with $C = \frac{6}{5 t_{in}} \, \left[ \rho_{turb} (t_{in}) \right]^{- \frac{5}{6}}$ and where $t_{in}$ is an initial time, from which the universe starts its development onwards. $C_0$ is an  integration constant, which can be determined from the initial condition $t = t_{in}$ and $\gamma_{turb} = 1 + \omega_{turb}$. Then, use of Eqs.~(\ref{it14}) and (\ref{it15}) yields
				\begin{equation}\label{it17}
					\omega(\phi) = \rho_{turb} \, + \, p_{turb} = \gamma_{turb} \rho_{turb} =$$
					$$= \gamma_{turb} \, e^{- 3 \gamma_{turb} H_0 \phi} \, \left[ \frac{C}{3 \gamma_{turb}} \left( e^{- \frac{5}{2} \gamma_{turb} H_0 t_{in}} - e^{- \frac{5}{2} \gamma_{turb} H_0 \phi} \right) \, + \, \frac{2 C_0 a_0^{\frac{5}{2} \gamma_{turb} H_0}}{5 \gamma_{turb}} \right]^{- \frac{6}{5}}
				\end{equation}
			and
				\begin{equation}\label{it18}
					V(\phi) = \rho_{turb} \, - \, p_{turb} = (1 - \omega_{turb}) \rho_{turb} =$$
					$$= (1 - \omega_{turb}) \, e^{- 3 \gamma_{turb} H_0 \phi} \, \left[ \frac{C}{3 \gamma_{turb}} \left( e^{- \frac{5}{2} \gamma_{turb} H_0 t_{in}} - e^{- \frac{5}{2} \gamma_{turb} H_0 \phi} \right) \, + \, \frac{2 C_0 a_0^{\frac{5}{2} \gamma_{turb} H_0}}{5 \gamma_{turb}} \right]^{- \frac{6}{5}}.
				\end{equation}
			As a consequence, considering the action (\ref{it1}), with Eqs.~(\ref{it17}) and (\ref{it18}), together with the assumption of being $\phi = t$, is just equivalent to consider isotropic turbulence in a de Sitter universe.
			
		\subsection{Effective quintessence-like power-law expansion}
			
			In this subsection we discuss the situation when the scale factor is given by $a(t) = a_0 t^{h_0}$, with $a_0$ and $h_0$ constant. For this case, the energy density of the turbulent part is
				\begin{equation}\label{it19}
					\rho_{turb} = t^{- 3 \gamma_{turb} h_0} \, \left[ \frac{5 C}{6 \left( 1 - \frac{5}{2} \gamma_{turb} h_0 \right)} \left( t^{1 - \frac{5}{2} \gamma_{turb} h_0} - t_{in}^{1 - \frac{5}{2} \gamma_{turb} h_0} \right) \, + \, C_0 a_0^{\frac{5}{2} \gamma_{turb} h_0} \right]^{- \frac{6}{5}},
				\end{equation}	
			being $C = \frac{6}{5 t_{in}} \, \left[ \rho_{turb} (t_{in}) \right]^{- \frac{5}{6}}$ and $t_{in}$, as before,  an initial time from which the universe starts its development onwards. Again, $C_0$ is a constant of integration, to be determined from the initial condition $t = t_{in}$ and $\gamma_{turb} = 1 + \omega_{turb}$. Repeating the procedure of the previous subsection, we can write
				\begin{equation}\label{it20}
					\omega(\phi) = \rho_{turb} \, + \, p_{turb} = \gamma_{turb} \rho_{turb} =$$
					$$= \gamma_{turb} \, t^{- 3 \gamma_{turb} h_0} \, \left[ \frac{5 C}{6 \left( 1 - \frac{5}{2} \gamma_{turb} h_0 \right)} \left( t^{1 - \frac{5}{2} \gamma_{turb} h_0} - t_{in}^{1 - \frac{5}{2} \gamma_{turb} h_0} \right) \, + \, C_0 a_0^{\frac{5}{2} \gamma_{turb} h_0} \right]^{- \frac{6}{5}}
				\end{equation}
		and		\begin{equation}\label{it21}
					V(\phi) = \rho_{turb} \, - \, p_{turb} = (1 - \omega_{turb}) \rho_{turb} =$$
					$$= (1 - \omega_{turb}) \, t^{- 3 \gamma_{turb} h_0} \, \left[ \frac{5 C}{6 \left( 1 - \frac{5}{2} \gamma_{turb} h_0 \right)} \left( t^{1 - \frac{5}{2} \gamma_{turb} h_0} - t_{in}^{1 - \frac{5}{2} \gamma_{turb} h_0} \right) \, + \, C_0 a_0^{\frac{5}{2} \gamma_{turb} h_0} \right]^{- \frac{6}{5}}.
				\end{equation}	
			Also, as remarked above, the scalar-tensor gravity given by the action (\ref{it1}), with Eqs.~(\ref{it20}) and (\ref{it21}), together with the assumption $\phi = t$, is here equivalent to consider isotropic turbulence in the dark energy component, now in an effective quintessence-like power-like expanding universe.
			
		\subsection{Phantom-like power law expansion}\label{subs}
			
			Here we will deal with a phantom-like power law expansion, given by $a(t) = a_0 (t_s - t)^{-h_0}$, which is known to give rise to Type I finite-time future singularities, with constant $a_0$ and $h_0$. The energy density of the turbulent part is
				\begin{equation}\label{it22}
					\rho_{turb} = (t_s - t)^{3 \gamma_{turb} h_0} \, \left\{ \frac{5 C}{6 \left( 1 + \frac{5}{2} \gamma_{turb} h_0 \right)} \left[ (t_s - t)^{1 + \frac{5}{2} \gamma_{turb} h_0} - (t_s - t_{in})^{1 + \frac{5}{2} \gamma_{turb} h_0} \right] \, + \, C_0 a_0^{\frac{5}{2} \gamma_{turb} h_0} \right\}^{- \frac{6}{5}},
				\end{equation}	
			where $C = \frac{6}{5 t_{in}} \, \left[ \rho_{turb} (t_{in}) \right]^{- \frac{5}{6}}$ and $t_{in}$ and  $C_0$ have exactly the same meaning as before. In this case
				\begin{equation}\label{it23}
					\omega(\phi) = \rho_{turb} \, + \, p_{turb} = \gamma_{turb} \rho_{turb} =$$
					$$= \gamma_{turb} \, (t_s - t)^{3 \gamma_{turb} h_0} \, \left\{ \frac{5 C}{6 \left( 1 + \frac{5}{2} \gamma_{turb} h_0 \right)} \left[ (t_s - t)^{1 + \frac{5}{2} \gamma_{turb} h_0} - (t_s - t_{in})^{1 + \frac{5}{2} \gamma_{turb} h_0} \right] \, + \, C_0 a_0^{\frac{5}{2} \gamma_{turb} h_0} \right\}^{- \frac{6}{5}}
				\end{equation}
	and			\begin{equation}\label{it24}
					V(\phi) = \rho_{turb} \, - \, p_{turb} = (1 - \omega_{turb}) \rho_{turb} =$$
					$$= (1 - \omega_{turb}) \, (t_s - t)^{3 \gamma_{turb} h_0} \, \left\{ \frac{5 C}{6 \left( 1 + \frac{5}{2} \gamma_{turb} h_0 \right)} \left[ (t_s - t)^{1 + \frac{5}{2} \gamma_{turb} h_0} - (t_s - t_{in})^{1 + \frac{5}{2} \gamma_{turb} h_0} \right] \, + \, C_0 a_0^{\frac{5}{2} \gamma_{turb} h_0} \right\}^{- \frac{6}{5}}.
				\end{equation}	
			Therefore, a scalar-tensor gravity with functions $\omega(\phi)$ and $V(\phi)$ given by  expressions (\ref{it23}) and (\ref{it24}), respectively, is equivalent to isotropic turbulence in a phantom-like power-like expanding universe.

			Thus, we have shown that an equivalence exists between isotropic turbulence and scalar-tensor gravity and that in the isotropic turbulence theory, finite-time future singularities appear too (see Sect.~\ref{subs}). As in previous sections, addition of a $R^2$ term to the Langrangian density can avoid the development of these finite-time future singularities \cite{N-O.2,A-N-O,C-L-N-O}.
			
\section{Conclusions}

In this paper, a reconstruction program has been dealt with which uses the freedom in the choice of arbitrary functions or potentials, for several models of modified gravity or scalar-tensor gravity, with the aim to reconstruct a background cosmology---quite complicated in general---which complies with the latest observational data. Along this line, a systematic search for different viable models of the dark energy universe, all of which give rise to finite-time, future singularities, has been undertaken, having as goal their detailed study to try to find common features, in the search for a general solution to this important problem. Specifically, it has been checked that the addition of an $R^2$ term provides indeed a universal tool to cure these finite-time future singularities.

More specifically, a universal procedure to cure all future singularities has been defined and carefully tested with the help of explicit examples, corresponding to each of the four different types of possible singularities, as classified in the literature. To start, the case of a fluid with an EoS which depends on a parameter $\alpha$, which can give rise to finite-time future singularities, has been considered. We have shown explicitly that, adding a specific function $G(H, \dot H, \ddot H, ...)$ to the EoS of the fluid, the different singularities can be cured, and it has been seen that this function can actually be considered as a modification of Einsteinian gravity.

The case of the non-minimal coupling of modified gravity to a matter Lagrangian has been investigated, too. The reconstruction scheme has been run for this case and the example of the cosmology given by the Hubble function $H(t) = h_s/(t_s - t)$ was analyzed. In Appendix A, the calculation has been done of the Friedmann equations for this non-minimal coupling of modified gravity to a matter Lagrangian. Further, the case of non-local gravity has been discussed. The example of the de Sitter space has been reproduced in the framework of non-local gravity, having been pointed out that such kind of theories can also give rise to finite-time, future singularities. Finally, the case of isotropic turbulence in the dark fluid universe was discussed, with the conclusion that the contribution of the turbulent part of dark energy can indeed be reproduced through the use of a scalar-tensor theory. As already indicated, several examples, corresponding to the different cases, have been studied in the paper in detail.

Concerning future perspectives, it is rather clear that, in order to address the singularity issue in all rigor it will be necessary to develop a fully-fledged theory of quantum gravity, what has proven to be a very difficult, up to now impossible, task. In any case, the presence of a finite-time, future singularity may cause various problems of physical nature, as instabilities in current black hole and stellar astrophysics. And, even without the recourse to a quantum theory of gravity, it is still meaningful to try to find natural scenarios, already at the classical level, that may cure this possible finite-time, future singularities. This has been successfully addressed in the paper, with the explicit construction of a general, universal procedure to cure all future singularities---what has been carefully tested with the help of specific examples corresponding to each one of the four different types of possible singularities, as classified in the literature.

To conclude, we should mention that quantum gravity effects (which usually contain different powers of the curvature) may become very important near future singularities. Even if classical considerations are, in principle, not valid there, it is known that the $\Box R$ term works against the singularity. Thus, an $R^2$ term (which will on its turn generate a $\Box R$ term) would, in principle, cure the possible singularities that could arise in the quantum theory, too. As a consequence, our method here could presumably be also extended to the quantum case without much trouble.
\medskip 

\noindent {\bf  Acknowledgments}

			We thank Sergei Odintsov for very useful discussions and advice. AJLR acknowledges a JAE fellowship from CSIC. The work has been partly supported by MICINN (Spain), projects FIS2006-02842 and FIS2010-15640, by the CPAN Consolider Ingenio Project, and by AGAUR (Generalitat de Ca\-ta\-lu\-nya), contract 2009SGR-994.		

	\appendix

		\section{Friedmann equations for the Lagrangian density (\ref{0})}\label{appendixA}
			
			In this appendix we carry out, in some detail, the calculations which are necessary to obtain the Friedmann equations for the Lagrangian density of Eq.~(\ref{0}).  Assuming a spatially-flat FRW universe and taking into account Eqs.~(\ref{6}) and (\ref{7}), Eq.~$(t,t)$ from Eq.~(\ref{1}) reads
			\begin{equation}\label{8}
				\frac{3 H(t)^2}{\kappa^2} + 3 C^2 \left( \dot H(t) + H(t)^2 \right) \frac{f'(R)}{a(t)^6 f(R)^2} - 3 C^2 H(t) \partial_t \left( \frac{f'(R)}{a(t)^6 f(R)^2} \right) + \frac{1}{2} \frac{C^2}{a(t)^6 f(R)} = 0,
			\end{equation}
		and Eq.~$(i,i)$ from Eq.~(\ref{1}),
			\begin{equation}\label{9}
				\frac{2 \dot H(t) + 3 H(t)^2}{\kappa^2} + C^2 \left( \dot H(t) + 3 H(t)^2 \right) \frac{f'(R)}{a(t)^6 f(R)^2} - 2 C^2 H(t) \partial_t \left( \frac{f'(R)}{a(t)^6 f(R)^2} \right) -$$
				$$- C^2 \partial_t \partial_t \left( \frac{f'(R)}{a(t)^6 f(R)^2} \right) - \frac{1}{2} \frac{C^2}{a(t)^6 f(R)} = 0.
			\end{equation}
		Taking derivatives with respect to time, one gets
			\begin{equation}\label{10}
				\partial_t \left( \frac{f'(R)}{a(t)^6 f(R)^2} \right) = \frac{1}{a(t)^6 f(R)^2} \left( f''(R) \dot R - 6 f'(R) H(t) - \frac{2 f'(R)^2 \dot R}{f(R)} \right)
			\end{equation}
	and		\begin{equation}\label{11}
				\partial_t \partial_t \left( \frac{f'(R)}{a(t)^6 f(R)^2} \right) = \frac{1}{a(t)^6 f(R)^2} \left[ 6 f'(R) \left( 6 H(t)^2 - \dot H(t) \right) + f''(R) \left( \ddot R - 12 \dot R H(t) \right) + \right.$$
				$$\left. + f'''(R) \dot R^2 + 2 f'(R)^2 \left( \frac{12 \dot R H(t) - \ddot R}{f(R)} \right) + 6 f'(R)^3 \frac{\dot R^2}{f(R)^2} - 6 f'(R) f''(R) \frac{\dot R^2}{f(R)} \right].
			\end{equation}

		Introducing Eq.~(\ref{10}) into Eq.~(\ref{8}) and taking into account that $$a(t) = a_0 \, \exp \left[ \int \limits_{t_0}^{t} H(t') dt' \right]$$  yields
			\begin{equation}\label{12}
				\frac{1}{6} f(R) \, + \, \frac{a_0^6 H(t)^2 \exp \left[ 6 \int \limits_{t_0}^{t} H(t') dt' \right]}{\kappa^2 C^2} f(R)^2 \, + \, \left( \dot H(t) \, + \, 7 H(t)^2 \right) f'(R) \, +$$
				$$+ \, \frac{2 H(t) \dot R}{f(R)} f'(R)^2 \, - \, \dot R H(t) f''(R) = 0,
			\end{equation}
		and putting Eqs.~(\ref{10}) and (\ref{11}) into Eq.~(\ref{9}), this reduces to
			\begin{equation}\label{13}
				- \frac{1}{2} f(R) \, + \, \frac{2 \dot H(t) + 3 H(t)^2}{\kappa^2 C^2} a_0^6 \exp \left[ 6 \int \limits_{t_0}^{t} H(t') dt' \right] f(R)^2 \, + \, 7 \left( \dot H(t) - 3 H(t)^2 \right) f'(R) \, +$$
				$$+ \, 2 \left( \frac{\ddot R - 10 \dot R H(t)}{f(R)} \right) f'(R)^2 \, - \, 6 \frac{\dot R^2}{f(R)^2} f'(R)^3 \, + \, \left( 10 \dot R H(t) - \ddot R \right) f''(R) \, +$$
				$$+ \, 6 \frac{\dot R^2}{f(R)} f'(R) f''(R) - \dot R^2 f'''(R) = 0.
			\end{equation}
		The other possibilities for Eq.~(\ref{1}) are identities.

		We also know that $R = 6 \dot H(t) + 12 H(t)^2$, which could be solved in terms of $t$ as $t = t(R)$. Taking this into account, Eq.~(\ref{12}) can be written as
			\begin{equation}\label{14}
				\frac{1}{6} f(R) \, + \, \frac{a_0^6 H(t(R))^2 \exp \left[ 6 \int \limits_{t_0}^{t(R)} H(t') dt' \right]}{\kappa^2 C^2} f(R)^2 \, + \, \left[ \dot H(t(R)) \, + \, 7 H(t(R))^2 \right] \frac{df(R)}{dR} \, +$$
				$$+ \, \frac{12 H(t(R)) \left[ \ddot H(t(R)) + 4 \dot H(t(R)) H(t(R)) \right] }{f(R)} \left( \frac{df(R)}{dR} \right)^2 \, -$$
				$$- \, 6 \left[ \ddot H(t(R)) + 4 H(t(R)) \dot H(t(R)) \right] H(t(R)) \frac{d^2 f(R)}{dR^2} = 0,
			\end{equation}
		and Eq.~(\ref{13}) as
			\begin{equation}\label{15}
				- \frac{1}{2} f(R) \, + \, \frac{2 \dot H(t(R)) + 3 H(t(R))^2}{\kappa^2 C^2} a_0^6 \exp \left[ 6 \int \limits_{t_0}^{t(R)} H(t') dt' \right] f(R)^2 \, + \, 7 \left[ \dot H(t(R)) - 3 H(t(R))^2 \right] \frac{df(R)}{dR} \, +$$
				$$+ \, 12 \, \frac{- 40 H(t(R))^2 \dot H(t(R)) + 4 \dot H(t(R))^2 - 6 H(t(R)) \ddot H(t(R)) + \dddot H(t(R))}{f(R)} \, \left( \frac{df(R)}{dR} \right)^2 \, -$$
				$$- \, 216 \frac{\left[ \ddot H(t(R)) + 4 H(t(R)) \dot H(t(R)) \right]^2}{f(R)^2} \left( \frac{df(R)}{dR} \right)^3 \, -$$
				$$- \, 6 \left[ - 40 H(t(R))^2 \dot H(t(R)) + 4 \dot H(t(R))^2 - 6 H(t(R)) \ddot H(t(R)) + \dddot H(t(R)) \right] \frac{d^2 f(R)}{dR^2} \, +$$
				$$+ \, 216 \frac{\left[ \ddot H(t(R)) + 4 H(t(R)) \dot H(t(R)) \right]^2}{f(R)} \frac{df(R)}{dR} \frac{d^2 f(R)}{dR^2} \, - \, 36 \left[ \ddot H(t(R)) + 4 H(t(R)) \dot H(t(R)) \right]^2 \frac{d^3 f(R)}{dR^3} = 0.
			\end{equation}
		Eqs.~(\ref{14}) and (\ref{15}) are the Friedmann equations for the Lagrangian density given by (\ref{0}) and constitute the two differential equations we were looking for $f(R)$.

\end{document}